\documentclass[aps,prl,twocolumn,showpacs,floatfix]{revtex4}

\usepackage{graphicx}
\usepackage{dcolumn}
\usepackage{bm}
\usepackage{times}
\usepackage{color}

\begin{document}
\bibliographystyle{apsrev}
\title{Resistivity plateau and extremely large magnetoresistance in NbAs$_2$ and TaAs$_2$}
\author{Yi-Yan Wang}
\author{Qiao-He Yu}
\author{Tian-Long Xia}\email{tlxia@ruc.edu.cn}
\affiliation{Department of Physics, Beijing Key Laboratory of Opto-electronic Functional Materials $\&$ Micro-nano Devices, Renmin University of China, Beijing 100872, P. R. China}

\date{\today}
\pacs{75.47.-m, 71.30.+h, 72.15.Eb}

\begin{abstract}
In topological insulators (TIs), metallic surface conductance saturates the insulating bulk resistance with decreasing temperature, resulting in resistivity plateau at low temperatures as a transport signature originating from metallic surface modes protected by time reversal symmetry (TRS). Such characteristic has been found in several materials including Bi$_{2}$Te$_{2}$Se, SmB$_{6}$ \emph{etc}. Recently, similar behavior has been observed in metallic compound LaSb, accompanying an extremely large magetoresistance (XMR). Shubnikov-de Hass (SdH) oscillation at low temperatures further confirms the metallic behavior of plateau region under magnetic fields. LaSb\cite{tafti2015resistivity} has been proposed by the authors as a possible topological semimetal (TSM), while negative magnetoresistance is absent at this moment. Here, high quality single crystals of NbAs$_2$/TaAs$_2$ with inversion symmetry have been grown and the resistivity under magnetic field is systematically investigated. Both of them exhibit metallic behavior under zero magnetic field, and a metal-to-insulator transition occurs when a nonzero magnetic field is applied, resulting in XMR (1.0$\times$10$^{5}\%$ for NbAs$_2$ and 7.3$\times$10$^{5}\%$ for TaAs$_2$ at 2.5 K $\&$ 14 T). With temperature decreased, a resistivity plateau emerges after the insulator-like regime and SdH oscillation has also been observed in NbAs$_2$ and TaAs$_2$.
\end{abstract}

\maketitle
\setlength{\parindent}{1em}

\section{Introduction}

\indent The magnetoresistance (MR) of materials is an interesting research topic in condensed matter community. Since the discovery of giant magnetoresistance (GMR) in magnetic multilayer\cite{baibich1988giant,binasch1989enhanced} and colossal magnetoresistance (CMR) in magnetic oxide materials\cite{moritomo1996giant,ramirez1997colossal}, people have always been looking for materials with higher MR. In general, nonmagnetic metals only have small MR. Recently, XMR of about $10^{5}\%-10^{6}\%$ has been observed in several nonmagnetic metals, such as TX (T=Ta/Nb, X=As/P)\cite{PhysRevX.5.011029,PhysRevX.5.031023,Ghimire2015Magneto,yang2015chiral,shekhar2015large,hu2015pi,PhysRevB.92.041203,borrmann2015extremely,wang2015helicity}, NbSb$_2$\cite{wang2014anisotropic}, LaSb\cite{tafti2015resistivity}, Cd$_3$As$_2$\cite{liang2015ultrahigh,he2014quantum} and WTe$_2$\cite{ali2014large,zhu2015quantum} et al. In these materials, TX is an important system which is composed of VA elements and pnicogen, all of which are proved to be Weyl semimetals\cite{xu2015discoveryTaAs,huang2015weyl,PhysRevX.5.031013DingHongTaAs,NatPhysDingHongTaAs,yang2015weyl,sun2015topological,xu2015discoveryNbAs,xu2015experimental,xu2015observation,liu2016evolution,di2015observation}. Extraordinary properties such as Fermi arc and chiral anomaly observed in TX have attracted great attention. The interesting properties of transition-metal monophosphides motivate us to research into other transition-metal phosphides with different crystal structures. TX$_2$ type compounds, which are made of VA elements and pnicogen, have a monoclinic structure with inversion symmetry. The antimonide NbSb$_2$\cite{wang2014anisotropic} has been investigated previously. In this work, we focus on the TX$_2$ type arsenides.

\indent In addition to XMR, the field induced resistivity plateau is also an interesting phenomenon. The plateau is a universal behavior in topological insulators\cite{kim2013surface,ren2010large}, which originates from the metallic surface conductance saturating the insulating bulk resistance\cite{tafti2015resistivity}. TRS protects the metallic surface state and is indispensable in TIs. Recent research in possible TSM candidate LaSb shows that there also exists plateau when TRS is broken\cite{tafti2015resistivity}. Although the specific mechanism of this feature is still a mystery, it is meaningful to find out more materials with such property.

\indent In this work, we have mainly studied the resistivity and magnetoresistance of two synthesized single crystalline compounds: NbAs$_2$ and TaAs$_2$. Metal-to-insulator transition and XMR arise at low temperature and high field. The observed resistivity plateau is similar to the universal behavior in TIs, in which TRS is preserved, while TRS is broken in our measurements as the case in LaSb. Analysis of SdH effect observed in NbAs$_2$ and TaAs$_2$ implies the complex Fermi surface structure and the two major oscillation frequencies reveal there may exist two Fermi pockets in each of the samples.

\begin{figure}[htbp]
\centering
\includegraphics[width=0.48\textwidth]{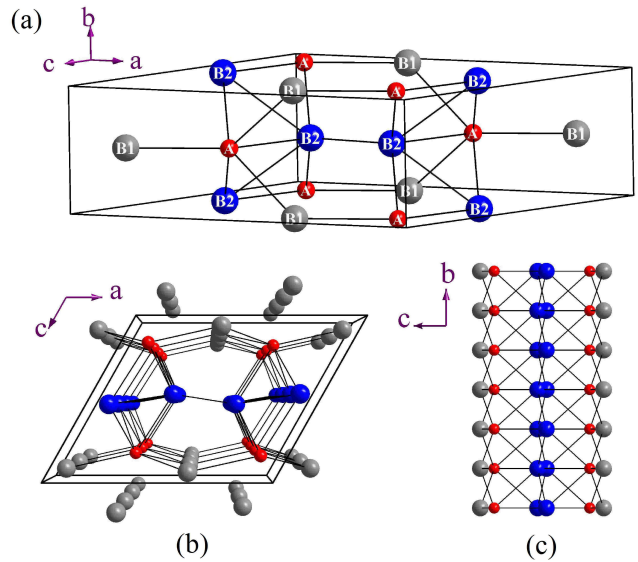}
\caption{(Color online) Crystal structure of NbAs$_2$ and TaAs$_2$. (a) Unit cell of the crystal. A represents Nb or Ta atoms, B1 and B2 represent type-I and type-II atoms of As, respectively. (b)-(c) Views of the structure from b-axis and a-axis directions, respectively.}\label{fig1}
\end{figure}

\section{experimental and crystal structure}

\indent High-quality single crystals of NbAs$_2$ and TaAs$_2$ were obtained by chemical vapor transport method. Firstly, polycrystal of these two compounds was prepared by solid state reaction. The mixtures of Nb/Ta powder and As powder were sealed in a quartz tube with a ratio of 1:2. For NbAs$_2$, the tube furnace was heated to 800$^{0}$C and held for 30h, for TaAs$_2$ the tube furnace was heated to 750$^{0}$C. Secondly, 5 mg/cm$^3$ of iodine was used as transport agent to grow single crystals, and the temperature gradient was set as 1050$^{0}$C-850$^{0}$C. All the obtained crystals grow more easily along b-axis and form rod-like crystals. The atomic proportions determined by energy dispersive X-ray spectroscopy (EDS, Oxford X-Max 50) were consistent with 1:2 for (Nb/Ta):As. X-ray diffraction (XRD) patterns were obtained from powder of single crystals on a Bruker D8 Advance X-ray diffractometer using Cu K$_{\alpha}$ radiation. TOPAS-4.2 was employed for the refinements. Resistivity measurements were performed on a Quantum Design physical property measurement system (QD PPMS-14T).

\begin{figure}[htbp]
\centering
\includegraphics[width=0.48\textwidth]{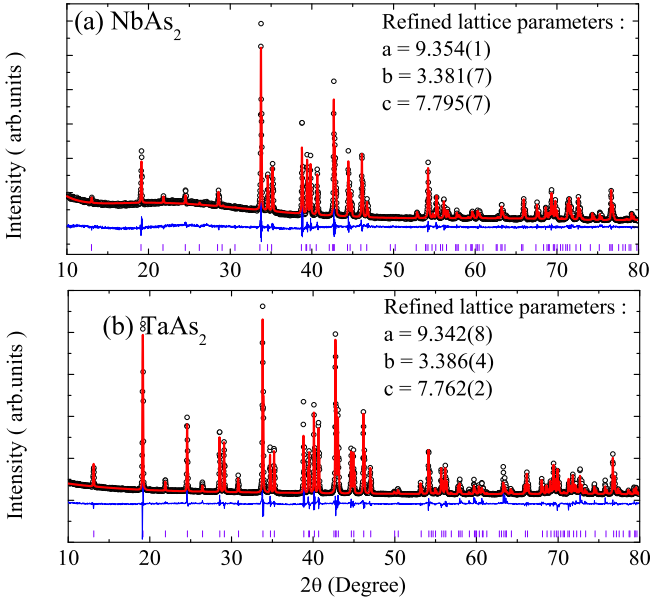}
\caption{(Color online) XRD patterns of NbAs$_2$ and TaAs$_2$ with refinement. Observed curves are in black circle and the calculated curves are in red line. The difference curve is in blue and the violet vertical lines denote the positions of Bragg reflections of NbAs$_2$ and TaAs$_2$ with PDF No.01-086-0520 and No.01-089-3409. Refined lattice parameters are shown in each picture. $R_{wp}$=6.69\%, 8.73\% for NbAs$_2$ and TaAs$_2$.}\label{fig2}
\end{figure}

\indent NbAs$_2$ and TaAs$_2$ crystallize in a complex structure\cite{bensch1995nbas2,furuseth1965arsenides}. Figure 1 shows the structure of these compounds from different views. In the structure, there is one type of Nb/Ta atoms denoted as A and two types of As atoms denoted as B1 or B2. Each A is surrounded by six B, including three B1 and three B2. Each B2 is coordinated to three other B2. In these crystals, b-axis is the easy growth axis, and this characteristic can be seen clearly from a-axis direction as shown in Figure 1(c). Figure. 2 (a) and (b) show powder X-ray diffraction patterns and refinements of NbAs$_2$ and TaAs$_2$ crystals. The reflections are well indexed in space group C$_{1 2/m1}$. The refined lattice parameters are given in the figures, consistent with previously reported  results\cite{bensch1995nbas2,furuseth1965arsenides,ling1981affinementstructureofTaAs2}.

\begin{figure}[htbp]
\centering
\includegraphics[width=0.48\textwidth]{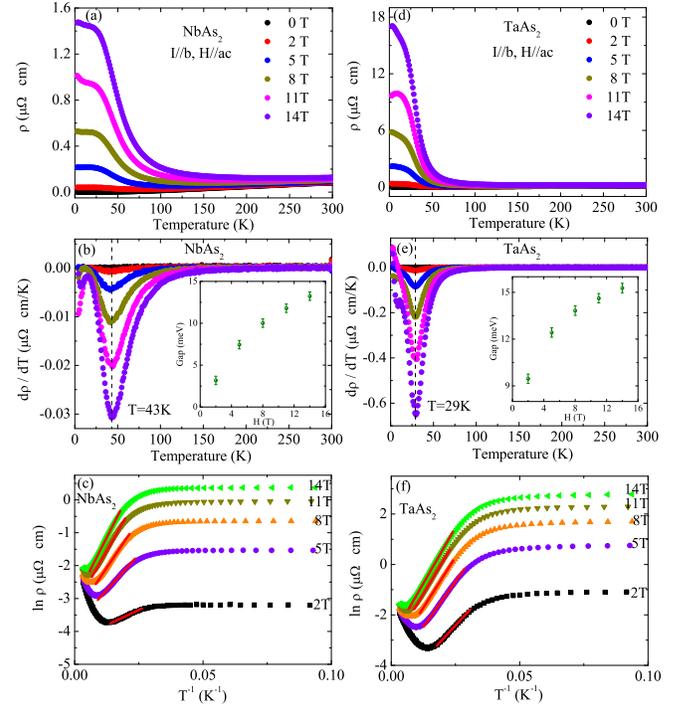}
\caption{(Color online) (a),(d) Resistivity of NbAs$_2$ and TaAs$_2$ plotted as a function of temperature under different magnetic field (H= 0, 2, 5, 8, 11, 14 T). I is parallel to b-axis and H is parallel to ac-plane. The metal-to-insulator-like transition and resistivity plateau are observed clearly. (b),(e) $d\rho/dT$ versus temperature for corresponding samples. The insets show energy gap of the insulator-like area. (c),(f) Plots of $ln\rho$ against the reciprocal of temperature $1/T$. The values of energy gap are obtained by fitting the insulator-like regions (the linear part in figures) using the relation $\rho (T)\propto exp(\xi/k_{\scriptscriptstyle B}T)$. The red lines indicate the regions used in fittings.}\label{fig3}
\end{figure}

\section{Results and discussions}

\indent Figure 3(a),(d) plot the temperature dependence of resistivity under different magnetic fields. The electric current is parallel to b-axis and the magnetic field is parallel to ac-plane. For both compounds, the temperature dependent resistivity at zero field exhibits a metallic behavior. The high residual resistivity ratio (RRR=75, 83 for NbAs$_2$ and TaAs$_2$, respectively) indicates high quality of the samples. Extremely large MR has been observed in both of them. At 2.5 K and 14 T, MR=$1.0\times10^5\%$ and $7.3\times10^5\%$ for NbAs$_2$ and TaAs$_2$, respectively. With temperature reduced, a magnetic field-induced metal-to-insulator-like transition is observed. An interesting feature is that a resistivity plateau appears instead of the resistivity continues to increase with further decreasing temperature. Temperatures of the resistivity plateau appearing are about 28 K and 19 K for NbAs$_2$ and TaAs$_2$, respectively. At high field, the plateau becomes distorted because of the effect of SdH oscillation, which will be discussed later. In general, topological insulators have insulating bulk state and metallic surface state. The TRS protects the metallic surface state in topological insulators. Along with the appearance of the metallic surface conduction, the insulating resistivity will reach saturation resulting in a resistivity plateau. In our situation, the existence of magnetic field breaks the TRS, but there is still a plateau at low temperature. More experiments are in great need to understand the intrinsic physics.

\begin{figure}[htbp]
\centering
\includegraphics[width=0.48\textwidth]{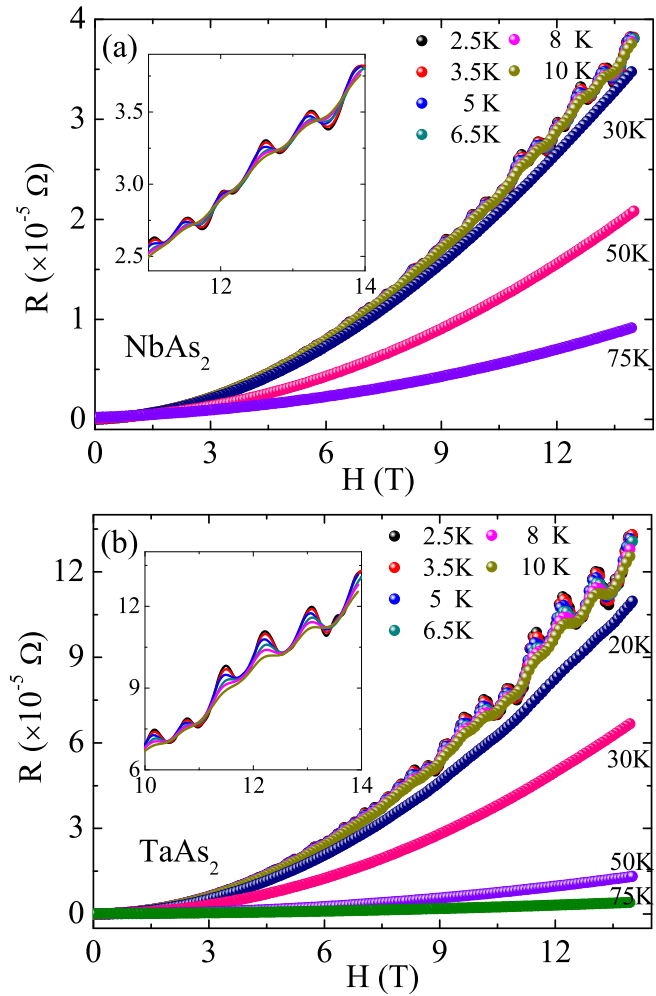}
\caption{(Color online) Magnetic field dependence of resistance of NbAs$_2$ (a) and TaAs$_2$ (b) from 2.5 K to 75 K. The insets show the enlarged part of R-H from 10-14 T, where the oscillation can be observed more clearly.}\label{fig4}
\end{figure}

\indent Plots of $d\rho/dT$ against temperature are displayed in Figure 3(b),(e). The temperature of metal-to-insulator transition is determined as $T_1$ where the value of $d\rho/dT$ becomes negative. From the figure, it can be judged that $T_1$ increases with the field. The starting temperature of the resistivity plateau ($T_2$) can be defined as the temperature at the minimum of $d\rho/dT$. As shown in the figure, $T_2$ remains unchanged with the increase of magnetic field. $T_1$ increases but $T_2$ keeps unchanged, which suggests that the range of the insulator-like area becomes larger along with increasing field, meanwhile the range of the resistivity plateau keeps invariant. The energy gaps are shown in the inset of Figure 5(b),(e). All the gap values are obtained through fitting the linear part in Figure 3(c),(f). At the magnetic field of H=14 T, the energy gaps are 13.2 meV and 15.3 meV for NbAs$_2$ and TaAs$_2$, respectively. As a whole, TaAs$_2$ has the larger gap than NbAs$_2$.

\begin{figure}[htbp]
\centering
\includegraphics[width=0.48\textwidth]{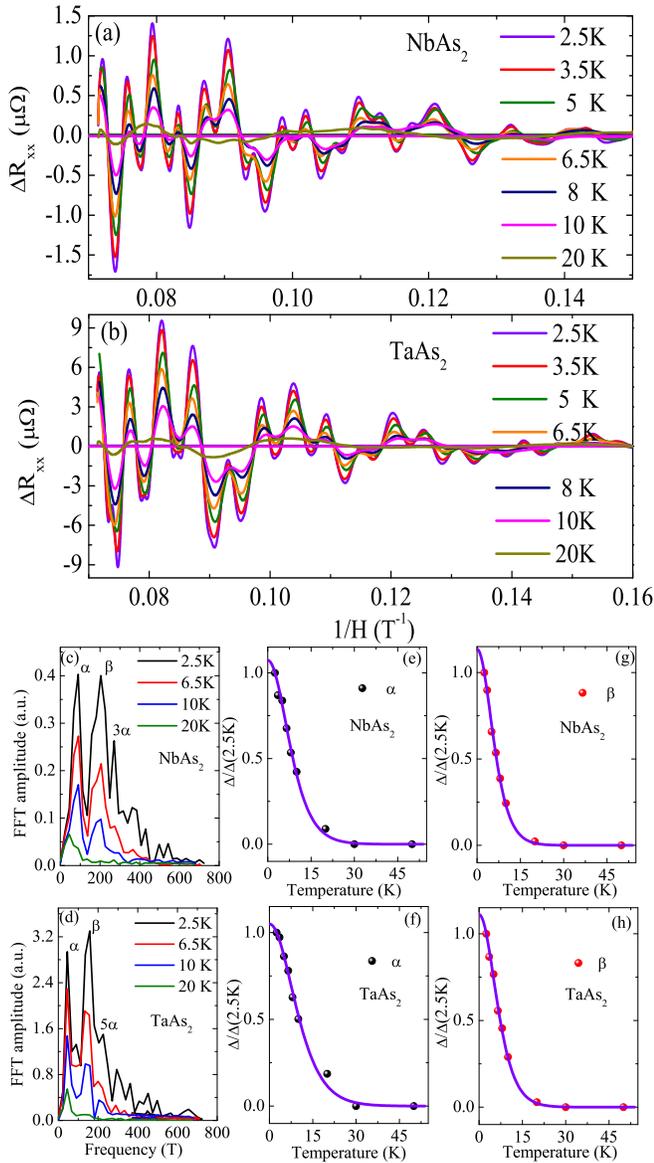}
\caption{(Color online) Analyses of oscillations in NbAs$_2$ and TaAs$_2$. (a),(b) The amplitude of SdH oscillations plotted as a function of the reciprocal of magnetic field. (c),(d) The FFT spectra of the corresponding SdH oscillations at 2.5 K,6.5 K,10 K and 20 K for NbAs$_2$ and TaAs$_2$. (e),(f) Temperature dependence of the relative FFT amplitude of oscillation frequency $\alpha$. The violet solid lines are the fitting results based on Lifshitz-Kosevitch formula, which yields the effective cyclotron resonant mass. (g),(h) The same processing of oscillation frequency $\beta$.}\label{fig5}
\end{figure}

\indent Figure 4(a) and (b) show the resistance of NbAs$_2$ and TaAs$_2$ as a function of field. Clear SdH oscillation was observed at low temperature and high field. The insets show the enlarged images of oscillating part. In both compounds, the MR-H curves (not shown here) all exhibit a semiclassical quadratic behavior ($MR\propto H^2$)\cite{abrikosov1988fundamentals}. With the increase of temperature, the MR becomes smaller and the oscillation gradually disappear.

\indent Figure 5(a) and (b) plot the oscillation amplitude $\Delta R_{xx}= R_{xx}-\langle R_{xx} \rangle$ of NbAs$_2$ and TaAs$_2$ against the reciprocal of magnetic field 1/H at various temperatures. The amplitude displays a complex periodic behavior and decreases with increasing temperature. There are several peaks in the fast Fourier transformation (FFT) spectra (Figure 5(c) and (d)), but the major peaks are $\alpha $ and $\beta $. The complexity of periodic behavior can be attributed to the effect of small peaks in the FFT spectra. For NbAs$_2$, the major oscillation frequencies are $F_{\alpha}$=90 T and $F_{\beta}$=204 T. For TaAs$_2$, $F_{\alpha}$=45 T and $F_{\beta}$=158 T. In SdH oscillation, the frequency $F$ is proportional to the cross sectional area $A$ of Fermi surface normal to the magnetic field and can be described using Onsager relation $F=(\phi_0/2\pi^2)A=(\hbar/2\pi e)A$\cite{shoenberg1984magnetic}. Two major peaks in FFT spectra imply there are two major Fermi pockets in NbAs$_2$ and TaAs$_2$, which is similar to the situation in NbSb$_2$\cite{wang2014anisotropic}. In Figure 5(e)-(h), we display the temperature dependence of the relative FFT amplitude of frequencies $\alpha$ and $\beta$ of NbAs$_2$ and TaAs$_2$, respectively. The thermal factor $R_T=(\lambda T)/sinh(\lambda T)$ in Lifshitz-Kosevitch formula\cite{shoenberg1984magnetic} has been employed to describe the temperature dependence of FFT amplitude $\Delta$. In the formula, $\lambda= (2\pi^2k_{\scriptscriptstyle B}m^*)/(\hbar e\bar{H})$, so the cyclotron effective mass $m^*$ can be obtained from the fitting result. All the data can be fitted well and yields the effective mass. For NbAs$_2$, $m^*_\alpha=0.20m_e$ and $m^*_\beta=0.27m_e$; for TaAs$_2$, $m^*_\alpha=0.17m_e$ and $m^*_\beta=0.24m_e$. ARPES experiments are needed to obtain more information about the Fermi surface and the collaborative work based on our samples are already in process\cite{ZhouSY}.

\section{Summary}

\indent In summary, single crystals of NbAs$_2$ and TaAs$_2$ have been grown successfully. Resistivity have been measured and magnetoresistance has been analyzed in detail. Field-induced metal-to-insulator transition and XMR are observed in both samples. The resistivity plateau similar to topological insulator emerges after the insulator-like regime, which is similar with previous work on LaSb crystal. Moreover, with the increase of field, clear SdH oscillation is observed in NbAs$_2$ and TaAs$_2$. The FFT spectra reveal there exist two major frequencies in the oscillation and the corresponding effective mass are obtained. As possible candidates of TSM, related ARPES studies on NbAs$_2$ and TaAs$_2$ are expected and further studies of the transport properties with angle dependence and resistivity under parallel field with current is in process.

\emph{Note added.} While this paper is being prepared, one related work on TaSb$_2$ is reported online\cite{TaSb2}, where resistivity plateau, XMR, SdH oscillation and negative MR have been observed by the authors.

\section{Acknowledgments}
\indent This work is supported by the National Natural Science Foundation of China (No.11574391), the Fundamental Research Funds for the Central Universities, and the Research Funds of Renmin University of China (No. 14XNLQ07).

\bibliography{bibtex}
\end{document}